\documentclass[aps,floats,showpacs,amssymb,notitlepage,tightenlines,amsmath12pt]{revtex4-1}

\usepackage{color}
\usepackage{amsmath}
\usepackage{amsfonts}
\usepackage{amscd}
\usepackage{epsfig}
\usepackage{amssymb}
\usepackage{tabularx}
\usepackage{longtable}
\usepackage{calligra}

\def\g{\gamma}

\def\l{\lambda}

\def\R{{\cal R}}

\def\p{\partial}

\def\B{{\cal B}}
\def\M{{\cal M}}

\DeclareMathOperator{\sech}{sech}

\begin{document}
\title{Strong cosmic censorship and Misner spacetime}

\author{Pedro Denaro and Gustavo Dotti}
\affiliation {Facultad de Matem\'atica, Astronom\'{\i}a y
F\'{\i}sica (FaMAF), Universidad Nacional de C\'ordoba and\\
Instituto de F\'{\i}sica Enrique Gaviola, CONICET.\\ Ciudad
Universitaria, (5000) C\'ordoba,\ Argentina}

\begin{abstract}
Misner spacetime is among the simplest solutions of Einstein's equation that exhibits a 
Cauchy horizon with a smooth extension beyond it. Besides violating strong cosmic censorship, 
this extension contains closed  timelike curves. We analyze 
the stability of the Cauchy horizon,  and prove that neighboring spacetimes  in 
one parameter families of solutions through Misner's 
in pure gravity, gravity coupled to a scalar field, or Einstein-Maxwell theory, 
end at the Cauchy horizon developing a  curvature singularity.
 \end{abstract}


\maketitle

\tableofcontents

\section{Introduction}
The possibility of smoothly extending a solution of Einstein's equations beyond the maximal Cauchy development 
of compact or asymptotically simple data is an undesirable feature of General Relativity (GR). From a 
a $3+1$ viewpoint,  the evolution of the three-metric ceases to be non unique at the Cauchy horizon, 
predictability being lost in a classical theory. 
Strong cosmic censorship (SCC) is the conjecture that this can only happen for  isolated, non generic solutions of GR. 
 Notably,  these pathologies occur among the  most important solutions 
of GR: all double horizon black holes in   the Kerr Newman family, those where either  charge, 
 angular momentum, or both, are nonzero. In these black holes the inner horizon is a Cauchy horizon 
for any  Cauchy surface connecting both copies of spatial infinity $i_o$, and 
the standard analytic extension beyond it is unique only if we enforce the non physical requirement of 
analyticity. In the rotating case, moreover, causality is completely lost in the  analytic extension, 
it being  possible to connect any two given events in this region with a future directed timelike curve  \cite{onil}; in particular, 
there are closed timelike curves -CTCs- through any point. A simple argument first given by Penrose 
in \cite{pen} (see also 
\cite{sp}) suggests that 
any perturbation of these solutions will actually end at the Cauchy horizon with a curvature singularity. 
The instability of the Cauchy horizon  was  illustrated for the Reissner Nordstr\"om  spacetime in \cite{Poisson:1990eh},
 using a model with a cross flow of outgoing and ingoing lightlike fluxes. An instability of transverse derivatives of test scalar fields 
along the  Cauchy horizon of extremal 
Reissner-Nordstr\"om black hole was recently found in \cite{Aretakis:2011ha,Aretakis:2011hc}, 
the analogous result for extremal Kerr black
holes is given 
in \cite{Lucietti:2012sf}.\\ 
Misner spacetime is obtained from the half $x^0 < x^1$ of Minkowski space by identifying points connected by 
a fixed boost. In spite of its simplicity, the resulting spacetime  has a rich structure that includes a Cauchy horizon with CTCs 
beyond it. Being a flat spacetime, it is possible to obtain explicit solutions for scalar and Maxwell test fields
and use these results in perturbation theory to order higher than one for the coupled scalar-gravity and Maxwell-gravity 
systems. We use these results, as well as perturbations in pure gravity, to 
show that Misner spacetime is an isolated solution in any of these theories. More precisely, 
we prove that given  a one parameter family of solutions  through Misner's in any of these theories, 
neighboring solutions develop  a curvature singularity that truncates the spacetime at the Cauchy horizon except for fine tuned cases.\\
We review the construction of Misner space in Section \ref{misner-sect} where we also analyze in detail the null geodesics, as they provide
insight in the evolution of massless fields. In Section \ref{scalar-sect} we prove that a zero scalar field on a Misner 
background is a non generic solution within the Einstein-scalar field theory: except for fine tuned cases, 
  perturbations of this solution  within the in Einstein-scalar field theory develop a curvature horizon that truncates the spacetime 
at the Cauchy horizon. The analogous result is proved for the Einstein-Maxwell theory in Section \ref{EM}, and then for pure 
gravity in Section \ref{grav}. \\
For simplicity, we have performed calculations in  compactified Misner space $\M_2 \times {\mathbb T}^2$, where $\M_2$ is two-dimensional 
Misner space, and ${\mathbb T}^2$ a 2-torus. We can recover the non-compact case by taking the limit $a, b \to \infty$ of the periods $a$
 and $b$  of the spatial 
coordinates $y$ and $z$. This amounts to a few changes in the test field expressions, such as replacing Fourier series in $(y,z)$
 with Fourier transforms.

\section{Misner spacetime} \label{misner-sect}

\begin{figure}[h] 
\centering
\includegraphics[width=130mm]{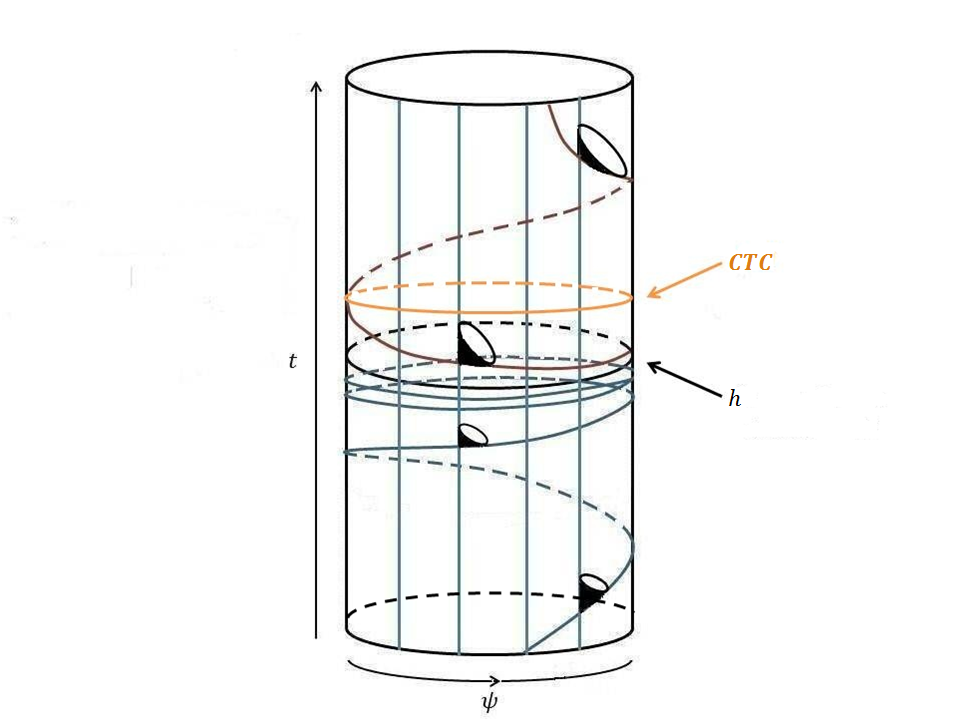}
\caption{Two dimensional Misner space:} \label{cili}
\end{figure}

Consider the half space $\widetilde \M _2$  of two-dimensional Minkowski spacetime 
\begin{equation}
ds^2= - (dx^0)^2 + (dx^1)^2 = - du \; dv, \;\; u=x^0 - x^1, \; v=x^0 + x^1,
\end{equation}
defined by the condition $v<0$. Introduce  coordinates
\begin{equation}\label{coords}
\psi= -\ln \left( \frac{v}{v_o} \right) , \; t = - u v,
\end{equation}
 $v_o<0$  a constant used for dimensional purposes. The line element in these coordinates is 
\begin{equation} \label{mm}
ds^2 = -d\psi dt - t d\psi^2, 
\end{equation}
and  the  boost 
\begin{equation} \label{boost1}
\B:(u,v) \to (\exp(\g) u, \exp(-\g) v), \g>0,
\end{equation}
is given by 
\begin{equation} \label{boost2}
(\psi,  t ) \to (\psi + \g,  t ).
\end{equation}
Two dimensional Misner space $\M_2$ is defined as the quotient of $\widetilde \M _2$ under the action of 
the subgroup $G= \{ \B^n | n \in {\mathbb Z} \}$ of the Lorentz group in $1+1$ dimensions, that is, 
points in $\widetilde \M _2$ which are  related by $\B ^n$ for some $n \in {\mathbb Z}$
 are considered equivalent, 
$\M_2$ being the set of equivalence classes. Since  $(\psi,  t )$ and $(\psi + n \g,  t ), n \in {\mathbb Z}$ 
represent the same 
point of $\M_2$, and  $ t $ extends from minus to plus infinity, two dimensional 
Misner space has the manifold structure of a cylinder
$S^1_{\psi}  \times {\mathbb R}_t$, $2\pi \psi/\g$ being an angular coordinate of $S^1$ (Figure \ref{cili}),
on which the flat Lorentzian metric (\ref{mm}) is defined.
Since the non vanishing vector field $\p/\p t$ is always
null, it gives a time orientation on $\M_2$: we define the future null half-cone as that where $\p/\p t$ belongs. 
This is  consistent with the 
time orientation $\p/\p x^0$ on the  covering  $\widetilde \M _2$, as can be   seen by lifting 
$\p/\p t$ to $\widetilde \M_2$, which gives $-(2v)^{-1} \; (\p/ \p x^0 - \p / \p x^1)$, a vector field that lies 
in the same half-cone of $\p / \p x^0$ since $v<0$ on $\widetilde \M _2$.

\subsection{Null geodesics}
\begin{figure}[h]
\centering
\includegraphics[width=130mm]{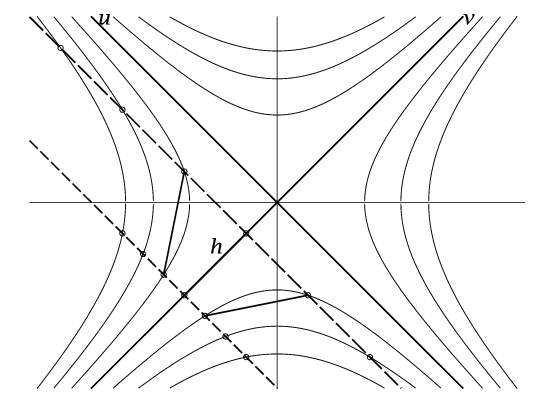}
\caption{Two dimensional Minkowski space: $x^0$ and $x^1$ are the vertical and horizontal axes. A few orbits of the Lorentz group are shown, 
including the $u$ and $v$ axes. Misner space $\M_2$ can be regarded as the strip between $\ell$ (null dashed-line) 
and $\B \ell$ (null line in long dashes) with points in $\ell$
identified with their image under $\B$ in $\B \ell$  (this is  the $t$-axis in Fig.\ref{cili}). Some of these 
pairs of identified points are marked with circles.  The three geodesic segments shown are,   
 from left to right,  time-like, null and space-like; they become   
 closed geodesics in $\M_2$. The closed   null  geodesic $h$ at $t=0$ ($u=0$) 
separates the non causal region $M_2^{>}$ above it ($t>0, u>0$) from the causal region $M_2^{<}$ below ($t<0, u<0$).}
\label{strip}
\end{figure}

The  image under $\B$ of  the Minkowskian null geodesic $v=v_o<0$ is the geodesic $v = \exp(-\g) v_o$; 
$\M_2$ can therefore be regarded  as the strip ${\cal S}_0 \subset \widetilde \M _2$ 
limited by $\ell = \{ (u,v_o), u \in {\mathbb R} \}$ and $\B \ell = 
\{ (u,\exp(-\g) \;v_o), u \in {\mathbb R} \}$, with the boundary points $(u,v_o)$ and $(\exp(\g) u, \
\exp(-\g) v_o)$  identified for every $u \in {\mathbb R}$. This construction  is  shown in Figure \ref{strip}, where some
 of the points to be identified
are marked with circles. 
A geodesic segment in $\widetilde \M _2$  connecting  identified points maps onto a  closed 
curve in $\M_2$ of square  length
\begin{equation}
\Delta s^2 = - \Delta u \Delta v = -2t (\cosh(\g) -1).
\end{equation}
These closed curves are timelike  
 in the $t>0$ sector and spacelike in the $t<0$ sector. The $t_o=0$ segment $h$ connecting $(u=0,v=v_o)$ with 
$(u=0,v=\exp(-\g) v_o)$ corresponds to  a closed null geodesic which is  a horizon separating the 
causally pathological   $t>0$ region from the globally hyperbolic $t<0$ region.
\\

The vector fields 
\begin{equation} \label{ns}
N_1 = \frac{\p}{\p t} , \;\; \tilde N_2 = -t \frac{\p}{\p t} + \frac{\p}{\p \psi} 
\end{equation}
are geodesic, null and, since $N_1^a N_2{}_b =-\tfrac{1}{2}$,  future oriented; this explains the arrangement
 of future half-cones in 
Figure \ref{cili}, from where it is readily seen that any future causal curve 
crossing $h$ (i.e., $\dot t \neq 0$ at $t=0$) must satisfy $\dot t > 0$ at $t=0$. Note  that 
 $N_1$ in  (\ref{ns}) is affine but $\tilde N_2$ is not. In fact, 
there is no globally defined affine geodesic field proportional to $\tilde N_2$.
It is however possible to  rescale separately $\tilde N_2$ in the $t>0$ and $t<0$ open sets
to obtain  a  future null affine geodesic field $N_2$ in each  of these regions:
\begin{equation}
N_2 = \begin{cases} - \frac{\p}{\p t} + \frac{1}{t} \frac{\p}{\p \psi} &, t > 0 \\
 \frac{\p}{\p t} - \frac{1}{t} \frac{\p}{\p \psi} &, t < 0 \end{cases}
\end{equation}
The integral curves of $N_1$ starting at $(t_0,\psi_0)$ at {\em affine parameter} $s=0$ 
are 
\begin{equation} \label{g1}
 t=t_0 + s, \; \psi=\psi_0, \;
 -\infty < s < \infty.
\end{equation}
These geodesics are complete and cross $h$.  
The integral curves of  $\tilde N _2$, starting at $(t_0,\psi_0)$ at {\em affine parameter} $s=0$ are 
the future incomplete null geodesics 
\begin{align}
&t=t_0 - s,  \psi=\psi_0 - \ln  \left( \frac{t_0-s}{t_0} \right),  -\infty < s < t_0 & \text{ if } t_0>0 \\
&t=0, \psi=-\ln \left(e^{-\psi_0}-\frac{s}{s_0} \right), -\infty<s< s_0 \; e^{-\psi_0}, \; s_0>0&  \text{ if } t_0=0  \label{c} \\
& t=t_0 + s, 
\psi=\psi_0 - \ln  \left( \frac{t_0+s}{t_0} \right),   -\infty < s < -t_0 & \text{ if } t_0<0 \label{t0<0}
\end{align}
For $t_0 \neq 0$ the above equations imply $\psi=\psi_0 - \ln \left( \frac{t}{t_0} \right)$, with $t \to 0^+$ ($t \to 0^-$) at 
the geodesic future end if $t_0>0$ ($t_0<0$).
These geodesic spiral,  asymptotically approaching $h$ as   
$s \to |t_0|^-$, see Figure \ref{cili}. 
This behavior can be understood using the construction in Figure \ref{strip}: a future directed segment along
 the $v$ direction,  starting at 
a point $p \in \ell$,  will reach $\B \ell$ at $q$ and emerge at the  equivalent point $q' \in \ell$ which
 lies closer to $h$ than $p$, and this process 
repeats indefinitely. 
The affine geodesic $h$ starting at  $(t_0=0, \psi_0)$ in the direction of $\tilde N_2$ (increasing $\psi$), 
has a tangent vector $(e^{\psi}/s_0)  \tfrac{\p}{\p \psi}$ (see (\ref{c})),   
which,  being non-periodic in $\psi$,  does {\em not} define a vector field on $h$.
This is a peculiar situation,  allowed to closed affine null geodesics in a 
Lorentzian geometry, for which the vector can scale in every turn and still have a constant norm.
 It  can be understood by noting  that $h$ lifts to the (future affine null)
  geodesic $\tilde h$ of $\widetilde M_2$ 
given by $u=0, v=v_o (e^{-\psi_0} - (s/s_0))$. The $n$-th turn of $h$ ($n=0,1,2,...$) 
lifts to the intersection $\tilde h_n$ of  $\tilde h$ with the strip 
 ${\cal S}_n \subset \widetilde \M _2$ 
limited by $\ell_n = \{ (u, \exp(-n \g)\; v_o), u \in {\mathbb R} \}$ and $\B \ell_n  = \ell_{n+1}$, 
The affine parameter span of this geodesic segment is  $\exp(-n\g)$ times 
that of the segment  $h_1$, yet the $\psi$ span is the same, therefore $\dot \psi$ scales as $\exp(+n\g)$ relative to the first 
turn.  
An extension of $M_2$ can be constructed that is geodesically complete, but it fails to be a 
Hausdorff manifold \cite{he}.\\

The  $\M_2^<$ subset defined by the condition $t<0$ is globally hyperbolic, 
any $t=$ constant surface is a Cauchy surface with Cauchy horizon $h$. The region 
$t>0$ violates causality in any possible form: 
given any two points $p$ and  $q$ in this  region, there is a 
future oriented timelike curve from $p$ to $q$ (these curves can  be easily constructed 
with the help of  Figure \ref{strip}.) Thus, $\M_2^<$ is a two-dimensional example of spacetime smoothly extensible beyond 
a Cauchy horizon, i.e., violating SCC, with the extension 
violating  causality in any possible form.\\

Four dimensional Misner spacetime $\M$ is the quotient of the $v<0$ half of 
Minkowski spacetime  $ds^2= - du \, dv + dy^2 + dz^2$ by the boosts (\ref{boost1}). The metric is 
\begin{equation} \label{mm4}
 \eta_{a b} \; dx^a dx^b = -d\psi dt - t d\psi^2 + dy^2+dz^2.
\end{equation}
As a Lorentzian manifold, $\M = \M_2 \times {\mathbb R}^2$ with the flat metric $dy^2 + dz^2$ on the ${\mathbb R}^2$ 
factor.
We may  consider $x$ and $y$ periodic with periods $a$ and $b$, that is, work instead with $\M_c = \M_2 \times 
{\mathbb T}^2$
with the flat metric on the torus. In any case, the open set defined by $t<0$ 
is globally hyperbolic, the Cauchy surfaces $t=$ (a negative) constant have the null hypersurface ${\cal H}$ defined by the 
condition $t=0$ as a future Cauchy horizon, 
and this four-dimensional spacetime violates strong form of the cosmic censorship conjecture, as it admits a smooth  extension beyond 
a Cauchy horizon. This is precisely 
what happens for the Kerr (also Kerr-Newman and Reissner-Nosrdtr\"om) black holes, with the inner horizon replacing ${\cal H}$. According to the 
 SCC  small departures 
in the initial data of these spacetimes will develop a globally hyperbolic spacetime with a null-like curvature singularity in place of the 
Cauchy horizon, and therefore 
admitting no  extension beyond it. \\

The purpose of this paper is to test SCC for $\M_<$, the $t<0$ sector of $\M$ (or its compact slice version $\M_c^<$ 
 with periodic $x$ and $y$) which, in spite of its simplicity (flat metric) has all possible pathologies 
 in 
an exact solution of Einsten's equation. In the following sections we will consider Misner spacetime as an exact solution 
in Einstein-scalar, Einstein-Maxwell and pure RG theories, and prove that in any of these theories it is an isolated 
solution.

\section{Instability of the Cauchy horizon in Einstein-scalar field theory} \label{scalar-sect}

Let $\Phi$ be a  massless scalar field minimally coupled to gravity  on the 
manifold $\M_c^<~=~S^1_{\psi} \times {\mathbb R}_{t<0} \times T^2_{(y,z)}$.
The field equations are
\begin{align} \label{Gs}
G_{ab} &= 8 \pi \; \left[ (\p_a \Phi) (\p_b  \Phi) - \tfrac{1}{2} g_{ab} \; g^{cd}(\p_c \Phi) (\p_d \Phi) \right] \\ 
0 &=  g^{c d}\; \nabla_d \nabla_c \Phi,  \label{Gs2}
\end{align}
where $\nabla_a$ is the Levi-Civita derivative of $g_{a b}$.
These equations  admit the solution
\begin{equation} \label{mis-sca}
\Phi_o=0, \; g_{a b}=\eta_{ab}.
\end{equation}
In this section we will prove that, although 
this solution  can be extended  to $\M_c = S^1_{\psi} \times {\mathbb R}_{t} \times T^2_{(y,z)}$ (i.e., add the $t \geq 0$ sector)  
any neighboring solution of the system (\ref{Gs})-(\ref{Gs2}) on the manifold 
 $\M_c^<$ develops a curvature singularity as
$t \to 0^-$.
To this end, consider a monoparametric family of solutions  $(\Phi_{\lambda}, \; (g_{\lambda}){}_{a b})$  such that  
\begin{equation} \label{ic}
(g_{\lambda=0}){}_{a b} = \eta_{ab}, \;\;\Phi_{\lambda=0}=\Phi_o = 0.
\end{equation}
Denote with   $n$ overdots 
 the $n-th$ derivative with respect to the parameter $\lambda$, evaluated at $\lambda=0$, then 
\begin{align}
\Phi_{\lambda} &=  \lambda \dot{\Phi} +\tfrac{1}{2} \lambda^2 \ddot{ \Phi} + ...\\
(g_{\lambda}){}_{a b} &= \eta_{ab} +  \lambda \dot{g}_{ab} +\tfrac{1}{2} \lambda^2 \ddot{g}_{ab} + ...
\end{align}
and similarly for any other tensor field. From  (\ref{Gs})-(\ref{Gs2}) we obtain for the Ricci scalar $\R$
\begin{equation} \label{Rsca}
\R_{\l} = \tfrac{1}{2} \lambda^2 \ddot{ \R} + {\cal O} (\l^3)
\end{equation}
where
\begin{equation} \label{Rsca2}
\ddot{\R} =16\pi\;  \eta^{ab} (\p_{a} \dot \Phi) (\p_{b} \dot \Phi),
\end{equation}
and 
\begin{equation} \label{Rsca3}
0 = \eta^{ab} \p_{a} \p_{b} \dot \Phi.
\end{equation}
Note that $\dot \Phi$ satisfies the equation of a  {\em test} scalar field on the Misner background (that is, without 
back-reaction effects). Yet,  it gives 
information on the Ricci scalar up to order two for the coupled scalar-gravity system (\ref{Gs})-(\ref{Gs2}). 
Motivated by this observation, we devote the following subsections to the study of  test scalar fields on $\M^<$, starting with $(y,z)$ independent fields, 
that is, scalar fields on $\M_2^<$. This will we used to prove that $\R$ in (\ref{Rsca}) diverges as $t \to 0^-$ for generic solutions of  
the theory  (\ref{Gs})-(\ref{Gs2}) near (\ref{ic}).\\

\subsection{Massless test scalar fields on $\M_2$}
Massless scalar fields $\tilde \Phi$ on $\widetilde \M_2$ satisfy $\partial_u \partial_v \tilde \Phi =0$, 
they are   a superposition $\tilde \Phi = R(u) + L(v)$ of left and right moving waves.
For  fields  defined on $M_2$, the extra condition
\begin{equation} \label{p1}
R(e^{\g} u) - R(u) = L(v) - L(e^{-\g}v) = c
\end{equation}
should be imposed, where $c$ is a constant and $c=0$ if $\lim_{u \to 0} R(u)$ exists.
Using the inverse of  (\ref{coords}), 
\begin{equation}\label{icoords}
v = v_o \; e^{-\psi}, \; \; 
u = \begin{cases} -v_o e^{\psi + \ln (|t|/v_o^2 ) } , & t>0\\
v_o e^{\psi + \ln (|t|/v_o^2)} , & t<0, 
\end{cases}
\end{equation}
and introducing $L(v)=L(v_o e^{-\psi})=:l(\psi)$, condition (\ref{p1}) reads $l(\psi)-l(\psi+ \g)=c$,
which implies that there exists a periodic function $\hat l$, $\hat l (\psi + \g) = \hat l (\psi)$, such that 
\begin{equation}
l(\psi) = \hat l (\psi) - \frac{c}{\g} \psi.
\end{equation}
A similar analysis for $R(u)$ in (\ref{p1}) using (\ref{icoords}) leads to
\begin{equation} \label{fi1}
\Phi =  \begin{cases} \hat l_{<} (\psi) +  \hat r_< (\psi + \ln (|t|/v_o^2)) + \frac{c_{<}}{\g} 
\; \ln (|t|/v_o^2), &  t < 0\\
\hat l_{>} (\psi) +  \hat r_>(\psi + \ln (|t|/v_o^2)) + \frac{c_{>}}{\g} 
\; \ln (|t|/v_o^2), & t>0
\end{cases}
\end{equation}
where all hatted functions are periodic with period $\g$, and therefore bounded if they are to be smooth in the 
corresponding $t>0$ or $t<0$ half-space. 
Note that  $\lim_{t \to 0} \Phi(t,\psi)$ along curves in the open set $\M_2^{<}$ 
cannot exist unless 
$c_<=0$ and  $\hat r_<$ is a constant, which may  then be absorbed into $\hat l_<$ to set $\hat r_< =0$
 (a similar analysis applies for  $\M_2^{>}$). 
Continuity across $h$ ($t=0$) would furthermore require $\hat l_< = \hat l _> =: \hat l$. Thus, 
  the only solutions that are continuous through $\M_2$ are the left moving waves 
$\Phi = \hat l (\psi)$ for all $t$. This is the condition, and the fields dealt with in \cite{groshev}. \\

\subsection{Scalar fields on $(\M_c^<, \eta_{ab})$}

The massless scalar field  $\Phi$ equation  on $(\M^<, \eta_{ab})$ is 
\begin{equation} \label{Fi4}
0= 4t\frac{\p^2 \Phi}{\p t^2}-4\frac{\p^2 \Phi}{\p \psi \p t}+4\frac{\p \Phi}{\p t}+\frac{\p^2 \Phi}{\p y^2}+
\frac{\p^2 \Phi}{\p z^2} = \Box_2 \Phi + \Delta_2 \Phi,
\end{equation}
where $ \Delta_2 = \p_y^2 + \p_z^2$ and $\Box_2$ is the massless scalar field operator  on $\M^<_2$. Solutions 
of (\ref{Fi4}) 
can be written as 
\begin{equation} \label{split}
\Phi = \phi_{(0)}(\psi,t) + \phi_{(1)}(\psi,t,y,z), 
\end{equation}
where 
\begin{equation}\label{fi0}
\phi_{(0)}(\psi,t) \equiv  \frac{1}{ab} \; \int_0^a dy  \; \int_0^b  dz \; \Phi(\psi,t,y,z)
\end{equation}
is a $(y,z)$-independent solution of  $\Box_2 \phi_0 =0$,
\begin{equation}\label{fi0}
\phi_{(0)}(\psi,t)  =  \hat l (\psi) +  \hat r(\psi + \ln (|t|/v_o^2)) + \tfrac{c}{\g} \; \ln (|t|/v_o^2),
\end{equation}
and 
\begin{equation} \label{fi1}
 \phi_{(1)} =  \sum_{\substack{k,l,n=-\infty \\ (l,n)\ne(0,0)}}^{\infty}C_{(k,l,n)}(t)e^{2\pi ik\psi/\gamma}
e^{2\pi i ly/a}e^{2\pi inz/b},  
\end{equation}
with 
\begin{equation}
C_{(k,l,n)}(t) := 
\frac{1}{ \g a b} \int_0^{\g} d\psi \; \int_0^a dy \; \int_0^b dz \; \phi_1 e^{-2\pi ik\psi/\gamma}e^{-2\pi i ly/a}e^{-2\pi inz/b}.
\end{equation}
Equation (\ref{Fi4}) reduces to 
\begin{equation} \label{B1}
	t\frac{d^2 C_{(k,l,n)}}{dt^2}+\left(1-i \nu\right)\frac{dC_{(k,l,n)}}{dt}-
\left(\frac{m}{2}\right)^2C_{(k,l,n)}=0, 
\end{equation}
where
\begin{equation}
m \equiv 2 \pi \; \sqrt{\tfrac{l^2}{a^2}+\tfrac{n^2}{b^2}}, \;\; \; \nu \equiv \frac{2\pi  k}{\gamma}.
\end{equation}
Introducing 
\begin{equation}
x \equiv m \sqrt{-t} \in (0,\infty) \;\;\; C_{(k,l,n)} = e^{i \nu \ln(x)} \; D_{(k,l,n)},
\end{equation}
(\ref{B1}) gives  a Bessel equation of imaginary order for $D_{(k,l,n)}$:
\begin{equation} \label{b1}
	x^2 \frac{d^2 D_{(k,l,n)}}{dx^2}+x \frac{dD_{(k,l,n)}}{dx}+
(x^2+\nu^2) D_{(k,l,n)}=0, 
\end{equation}
which admits the following two real, bounded, linearly independent ${\mathbb C}^{\infty}$ solutions for 
$x~\in~(0,\infty)$ (we follow the notation and conventions in  \url{http://dlmf.nist.gov/10.24}~):
\begin{equation} \label{bes}
\widetilde J_{\nu}(x) := \sech \left( \tfrac{1}{2} \pi \nu \right) \text{{Re}} (J_{i \nu}(x)) \;\;\; 
\widetilde Y_{\nu}(x) := \sech \left( \tfrac{1}{2} \pi \nu \right) \text{{ Re}} (Y_{i \nu}(x)).
\end{equation}
Thus
\begin{equation}
C_{(k,l,n)} = \left( A_{(k,l,n)} \widetilde J_{\nu}(x) + B_{(k,l,n)} \widetilde Y_{\nu}(x) \right)   e^{i \nu \ln(x)}
\end{equation}
and $A_{(-k,-l,-n)}=A_{(k,l,n)}^*$ for real $\Phi$.
The functions (\ref{bes}) satisfy
\begin{equation}
\widetilde J_{\nu}(x)=\widetilde J_{-\nu}(x), \;\;\; \widetilde Y_{\nu}(x)=\widetilde Y_{-\nu}(x)
\end{equation}
and have the following asymptotic behavior: as $x \to \infty$ ($t \to -\infty$)
\begin{align}
\widetilde J_{\nu}(x) &= \sqrt{\tfrac{2}{\pi x}} \,\cos\left( x-\tfrac{\pi}{4}\right)  + {\cal O}(x^{-3/2})\\
\widetilde Y_{\nu}(x) &= \sqrt{\tfrac{2}{\pi x}} \, \sin\left( x-\tfrac{\pi}{4}\right) + {\cal O}(x^{-3/2}),
\end{align}
as $x \to 0^+$ ($t \to 0^-$)
\begin{align} \label{a1}
\widetilde J_{\nu}(x) &= \sqrt{\tfrac{2 \tanh(\pi \nu/2)}{\pi \nu}}  \,\cos\left( \nu \ln(x/2)-\gamma_{\nu}\right)  + {\cal O}(x^{2})\\  \label{a2}
\widetilde Y_{\nu}(x) &= \sqrt{\tfrac{2 \coth(\pi |\nu|/2)}{\pi |\nu|}}  \,\sin\left( |\nu| \ln(x/2)-\gamma_{|\nu|}\right)  + {\cal O}(x^{2})
\end{align}
where $\gamma_{\nu}$ is defined by
\begin{equation}
\exp(i \gamma_{\nu}) =  \left( \frac{\sinh(\pi\nu)}{\pi \nu} \right) ^{1/2}\Gamma(1+i \nu) 
\end{equation}

\subsection{Instability of the Cauchy horizon in $\M_c^<$}

\noindent 
The instability of the Cauchy horizon in $\M_c^<$ is expressed as follows: \\

\noindent
{\bf Theorem 1:} Let $((g_{\lambda})_{ab}, \Phi_{\lambda})$ be a one-parametric family of solutions for the
 Einstein-real scalar field 
equations (\ref{Gs})-(\ref{Gs2}) on the manifold  $\M_c^<~=~S^1_{\psi} \times {\mathbb R}_{t<0} \times T^2_{(y,z)}$. 
Assume that $\lambda=0$ corresponds to Misner spacetime (\ref{mis-sca}). Let $\R_{\l}$ be the Riccci scalar of $g_{\l}$, then 
equations (\ref{Rsca})-(\ref{Rsca3}) hold and,  generically, $\ddot{ \R} \sim 1/t$  as $t \to 0^-$.\\

\noindent
{\bf Proof: } There only remains to prove that, with the exception of fine tuned solutions, 
$\ddot{ \R} \sim 1/t$  as $t \to 0^-$. From (\ref{Rsca3}),
\begin{equation}
\ddot{ \R} = \ddot{ \R}_{(0)(0)}+ \ddot{ \R}_{(1)(1)}+2  \ddot{ \R}_{(0)(1)} 
\end{equation}
where 
\begin{equation}
\ddot{ \R}_{(i)(j)} = -4 \frac{\p \phi_{(i)}}{\p \psi} \frac{\p \phi_{(j)}}{\p t} -4 \frac{\p \phi_{(i)}}{\p t} \frac{\p \phi_{(j)}}{\p \psi} 
+8 \frac{\p \phi_{(i)}}{\p t} \frac{\p \phi_{(j)}}{\p t}+ 2  \frac{\p \phi_{(i)}}{\p y} \frac{\p \phi_{(j)}}{\p y} + 
 2  \frac{\p \phi_{(i)}}{\p z} \frac{\p \phi_{(j)}}{\p z} 
\end{equation}
and $\phi_{(i)}, i=0,1$ were defined in (\ref{split})-(\ref{fi1}). \\
From (\ref{fi0}) we obtain
\begin{equation} \label{divsca}
\ddot{ \R}_{(0)(0)} = \frac{4}{\gamma t} \left[  \hat r'(\psi + \ln (|t|/v_o^2)) + \frac{c}{\gamma} \right]
\left[ \frac{c}{\gamma} - \hat l'(\psi) \right]
\end{equation}
Since the derivative of a periodic function cannot be a nonzero constant, $\ddot{ \R}_{00}$ can only vanish 
if either $c$ and $\hat r$ vanish or $c$ and $\hat l$  vanish. For generic one-parametric solutions of the 
Einstein-scalar field theory $\ddot{ \R}_{00} \sim 1/t$ near the Cauchy horizon. This divergence could only be canceled 
by the $1/t$ contribution from $\ddot{ \R}_{11}$ (implied by the asymptotic behavior (\ref{a1})-(\ref{a2})) by
 fine tuning the constants in these independent pieces of the scalar field. \\
If we restrict to fields that decay along past directed causal curves, we need to set $c=0$. This does not prevent the 
divergence (\ref{divsca}) except, once again, for the fine tuned case of pure left or right moving waves.\\

\section{Instability of the Cauchy horizon in Einstein-Maxwell theory} \label{EM}

Let $F_{ab}$ be a Maxwell field coupled to gravity on the 
manifold $\M^<_c~=~S^1_{\psi} \times {\mathbb R}_{t<0} \times T^2_{(y,z)}$.
The Einstein-Maxwell field equations
\begin{align} \label{GM}
&R_{ab} = 2  F_{ac} F_{bd} g^{cd}- \tfrac{1}{2} g_{ab} F_{cd} F_{ef} g^{ec} g^{df}  \\ 
&\nabla_{[a}F_{bc]} =0 , \; \; \nabla^a F_{ab} =0, \label{GM2}
\end{align}
 admit the solution
\begin{equation} \label{mis-max}
F_{ab}=0, \; g_{a b}=\eta_{ab}, 
\end{equation}
which can be extended  to $\M_c = S^1_{\psi} \times {\mathbb R}_{t} \times T^2_{(y,z)}$. 
In this section we will prove that generic  neighboring solution of the system (\ref{GM})-(\ref{GM2}) on the manifold 
 $\M_c^<$ develop a curvature singularity as
$t \to 0^-$. The Ricci scalar vanishes identically for the Einstein-Maxwell system, 
the singularity arises  in the quadratic curvature invariant 
\begin{equation} \label{q}
{\cal Q} = R_{a b} R_{cd} g^{a c} g^{b d}.
\end{equation}
Consider a monoparametric family of solutions  $((F_{\lambda})_{a b}, \; (g_{\lambda}){}_{ab})$  such that  
\begin{equation} \label{ic2}
(g_{\lambda=0}){}_{a b} = \eta_{ab}, \;\;(F_{\lambda=0})_{ab}= 0.
\end{equation}
As in the previous Section,  $n$ overdots are used to indicate 
 the $n-th$ derivative with respect to the parameter $\lambda$, evaluated at $\lambda=0$. We have 
\begin{align}
(F_{\lambda}){}_{a b} &=  \lambda \dot{F}_{ab} +\tfrac{1}{2} \lambda^2 \ddot{ F}_{ab} + ...\\
(g_{\lambda}){}_{a b} &= \eta_{ab} +  \lambda \dot{g}_{ab} +\tfrac{1}{2} \lambda^2 \ddot{g}_{ab} + ...
\end{align}
and 
\begin{equation} \label{Qsca}
{\cal Q}_{\l} = \tfrac{1}{4!}\;  \lambda^4 \;  \ddddot{ {\cal Q}} + {\cal O} (\l^5)
\end{equation}
where
\begin{align} \label{Qsca4}
\ddddot{\cal Q} &= \ddot R_{a b} \ddot R_{cd} \eta^{a c} \eta^{b d}\\  \label{Qsca4b}
\ddot R_{a b} &=  2  \dot F_{ac} \dot F_{bd} \eta^{cd}- \tfrac{1}{2} \eta_{ab} \dot F_{cd} \dot F_{ef} \eta^{ec} \eta^{df}. 
\end{align}
Note that  $\dot F$ satisfies the equations of a test Maxwell field {\em on the Misner background}, that is 
\begin{equation} \label{fdot}
\nabla_{[a} \dot F_{bc]} =0  ,  \; \; \eta^{bc} \nabla_c \dot F_{a b}=0,
\end{equation}
where $\nabla_c$ is the covariant derivative  of $\eta_{ab}$,
and that this  test Maxwell field gives information on the leading term of the curvature scalar ${\cal Q}$, 
which is fourth order in $\l$.

\subsection{Maxwell fields on $\M_c^<$}
Test Maxwell fields on the $M_c^<$ background are relevant to the Cauchy horizon stability problem  because, according to equations (\ref{Qsca4}) and (\ref{fdot}), 
they give the leading order contribution to the ${\cal Q}=R_{ab} R^{ab}$ curvature scalar in Einstein-Maxwell theory on this manifold. \\
The second Betti number of   $\M_c^<~=~S^1_{\psi} \times {\mathbb R}_{t<0} \times T^2_{(y,z)}$
is $3$, the three dimensional space of closed non-exact two forms is generated by $d \psi \wedge dy, 
d \psi \wedge dz$ and $dy \wedge dz$. Since these two-forms are divergence free for the flat metric $\eta$, the general 
solution of 
the Maxwell equations on the  background $\eta_{ab}$ is 
\begin{equation} \label{tmf}
F = K d \psi \wedge dy + L d \psi \wedge dz + M dy \wedge dz + d A^{(0)} + d A^{(1)}
\end{equation}
where, as done with the scalar field,  we have split $A_b = A^{(0)}_b + A^{(1)}_b$ with $ \pounds_{\p/\p y} A^{(0)}_b = \pounds_{\p/\p z} A^{(0)}_b =0$.
We  chose the one-forms $A_b^{(j)}$  in the Lorenz gauge $\nabla^b A_b^{(j)}=0$, then Maxwell equations 
reduce to    
\begin{equation} \label{max2}
\nabla^b A_b^{(j)}=0, \;\;\; \eta^{a b} 
\nabla_a \nabla_b A_c^{(j)}=0
\end{equation}
Introducing  $A^{(0)}_b(\psi,t) = \sum_{k \in {\mathbb Z}} C_b^k(t) \exp(2\pi i k \psi/\gamma)$   in (\ref{max2}) we find, 
after treating separately the $k=0$ and $k \neq 0$ terms and then  summing up the series, that 
\begin{align} \nonumber
	A^{(0)}_{\psi}(\psi,t) & =2at+\widehat{l}(\psi)+\widehat{r}(\psi+\ln(-t/v_o^2)) \\ \nonumber
	A^{(0)}_{t}(\psi,t) & =a+t^{-1}b+t^{-1}\widehat{r}(\psi+\ln(-t/v_o^2)) \\ \nonumber
	A^{(0)}_y(\psi,t) & = c_y\ln(-t/v_o^2)+\widehat{l}_y(\psi)+\widehat{r_y}(\psi+\ln(-t/v_o^2) \\ \nonumber
	A^{(0)}_z(\psi,t) & = c_z\ln(-t/v_o^2)+\widehat{l}_z(\psi)+\widehat{r_z}(\psi+\ln(-t/v_o^2)) 
\end{align}
This can be simplified using the residual gauge freedom $A^{(0)}_c \to A^{(0)}_c + \p_c \chi, \; \eta^{ab} \p_a \p_b \chi =0$.
Taking an appropriate  $\chi$ of the form (\ref{fi0}) we get a vector potential 
of the form
\begin{align} \nonumber
	A^{(0)}_{\psi}(\psi,t) & = 2a t+\widehat{l}_0 \\ \nonumber
	A^{(0)}_{t}(\psi,t) & = a \\ \nonumber
	A^{(0)}_y(\psi,t) & = c_y\ln(-t/v_o^2)+\widehat{l}_y(\psi)+\widehat{r_y}(\psi+\ln(-t/v_o^2)) \\
	A^{(0)}_z(\psi,t) & = c_z\ln(-t/v_o^2)+\widehat{l}_z(\psi)+\widehat{r_z}(\psi+\ln(-t/v_o^2)) \label{max3}
\end{align}
with $a, \hat l_0, c_y$ and $c_z$ constants (the irrelevant constant in $A^{(0)}_{t}(\psi,t)$ can be  gauged away  using the non-periodic 
 harmonic function $\chi_0=-a \psi$). For the Maxwell field we obtain
\begin{equation} \label{maxf}
\setlength{\arraycolsep}{1em}
F^{(0)} = d A^{(0)}= \left(  \begin{array}{rrcc}    0 & -2a & \widehat l_y{}'+ \widehat r_y{}' & \widehat l_z{}' + \widehat r_z{}'\\
 2a&0& (c_y +  \widehat r_y{}')/t&   (c_z +  \widehat r_z{}')/t \\
*&*&0 & 0\\
*&*&0&0 \end{array} \right)
\end{equation}
where we have omitted the arguments in the periodic (hatted) functions. 
The two field invariants for (\ref{maxf}) are 
\begin{equation} \label{inv1}
F^{(0)}_{ab} {F^{(0)}}^{ab} = -32 a^2 + 8 \; t^{-1} \;[c_y{}^2 + c_z{}^2 + c_y (\widehat r_y{}{}'- \widehat l_y{}')
+c_z(\widehat r_z{}'-\widehat{l}_z{}')-\widehat{l}_y{}'\;  \widehat r_y{}'-\widehat{l}_z{}' \;\widehat r_z{}'] 
\end{equation}
and
\begin{equation} \label{inv2}
\epsilon^{abcd} F^{(0)}_{ab} F^{(0)}_{cd} = -16 t^{-1} [ l'_z r'_y-l'_y r'_z + c_y (l'_z+r'_z) -c_z (l'_y+r'_y) ].
\end{equation}

\subsection{Instability of the Cauchy horizon in $\M_c^<$}
\noindent 
The instability of the Cauchy horizon of $\M_c^<$ in the Einstein-Maxwell theory is expressed in the following  \\

\noindent
{\bf Theorem 2:} Let $((g_{\lambda})_{ab}, (F_{\lambda})_{ab})$ be a one-parametric family of solutions for the
 Einstein-Maxwell field 
equations (\ref{GM})-(\ref{GM2}) on the manifold  $\M_c^<~=~S^1_{\psi} \times {\mathbb R}_{t<0} \times T^2_{(y,z)}$. 
Assume that $\lambda=0$ corresponds to Misner spacetime (\ref{mis-max}). Let ${\cal Q}_{\lambda}$ be the square Riccci scalar (\ref{q}) of $g_{\l}$, 
then equations (\ref{Qsca})-(\ref{fdot}) hold and,  generically, $\ddddot{ {\cal Q}}$ diverges at least as $\sim 1/t^2$  as $t \to 0^-$.\\

\noindent
{\bf Proof: } According to  equations (\ref{Qsca4}) and (\ref{Qsca4b}) $\ddddot{ {\cal Q}}$ is quartic on $\dot F_{ab}$. Since  $\dot F_{ab}$ 
satisfies Maxwell equations on the flat Misner background (see equation (\ref{fdot})), it is of the form (\ref{tmf}). 
 We will focus on 
the contribution $\ddddot{ {\cal Q}}'$ to $\ddddot{ {\cal Q}}$  that is  quartic in $F^{(0)} = d A^{(0)}$ in (\ref{tmf}).
 Note from (\ref{maxf}) that the general $F^{(0)}$ field is finite on any 
Cauchy slice in 
$\M_c^<$. A stronger condition of decay as $t \to -\infty$ can be enforced by requiring $a=0$ (see (\ref{inv1}) and (\ref{inv2})). 
In any case, the contribution $\ddddot{ {\cal Q}}'$, obtained by replacing $\dot F$ with (\ref{maxf}) in (\ref{Qsca4}) and (\ref{Qsca4b}) 
is
\begin{multline} \label{contr}
\ddddot{ {\cal Q}}' = 512 a^4 - 256 a^2 \; t^{-1} \;[c_y{}^2 + c_z{}^2 + c_y (\widehat r_y{}{}'- \widehat l_y{}')
+c_z(\widehat r_z{}'-\widehat{l}_z{}')-\widehat{l}_y{}'\;  \widehat r_y{}'-\widehat{l}_z{}' \;\widehat r_z{}'] \\
 + 16 \; t^{-2} \; [ \widehat{l}_z{}'{}^2 \; \widehat r_y{}'{}^2 + \widehat{l}_y{}'{}^2 \; \widehat r_z{}'{}^2 + 2 \widehat{l}_z{}'{}^2\; 
\widehat r_z{}'{}^2
+ 2 \widehat{l}_y{}'{}^2 \; \widehat r_y{}'{}^2 + 2 \widehat{l}_y{}' \;\widehat{l}_z{}' \; \widehat r_y{}'  \; \widehat r_z{}'  + ... 
+ 2c_y{}^4+ 2 c_z{}^4 ] 
\end{multline}
where the missing terms in the $t^{-2}$ coefficient involve growing powers of $c_y$ and $c_z$ times  derivatives of periodic functions.
As  $t \to 0^-$, (\ref{contr}) behaves as a bounded function times $t^{-2}$. 
This divergence 
could (in principle) be canceled out by the remaining contributions to  $\ddddot{ {\cal Q}}$, but this could  only be done 
by fine tuning, and will not be the case 
for generic mono-parametric solutions of the Einstein-Maxwell system.

\section{Instability of the Cauchy horizon in pure gravity} \label{grav}

The Cauchy horizon of $M_c^<$ can also be seen to be unstable in the context of pure gravity. Consider
a mono-parametric family  of Ricci flat metrics through $\eta_{ab}$  in (\ref{mm4}):
\begin{equation} \label{gl}
(g_{\lambda}){}_{a b} = \eta_{ab} +  \lambda \dot{g}_{ab} +\tfrac{1}{2} \lambda^2 \ddot{g}_{ab} + ...
\end{equation}
As is well known, any algebraic curvature scalar for a vacuum metric is a polynomial on $K:=R_{abcd}R^{abcd}$ and 
$L:=\epsilon_{a b p q}R^{pq}{}_{cd}R^{abcd}$. For  (\ref{gl}) we obtain
\begin{equation}
K_{\lambda} =  \lambda^2 \dot R_{abcd} \dot R_{efgh} \eta^{ae} \eta^{bf}\eta^{cg}\eta^{dh} + ...
\end{equation}
and similarly for $L_{\lambda}$, 
that is, knowledge of a linearized solution $\dot{g}_{ab}$ of Einstein's equation provides information on the dominant contributions 
to $K$ and $L$, which are  second order in $\lambda$.\\
Linear gravity on the background (\ref{mm4}) can be approached using the formalism in \cite{Ishibashi:2011ws}, 
which applies to warped metrics of any dimensions 
 with an Einstein compact  Riemannian manifold factor which, in our case, is the the trivial 2-torus flat metric $dy^2+dz^2$.
 Three different families of modes arise, 
tensor, vector and scalar,  which satisfy decoupled equations. Among them, the simplest contributions are  the two zero modes in the tensor sector, which 
are  constructed using 
the divergence free trace free harmonic symmetric tensors $dx \otimes dx - dy \otimes dy$ and $dx \otimes dy + dy \otimes dx$ on 
${\mathbb T}^2$. For these, the metric perturbation is 
\begin{equation} \label{tgp}
\dot g _{ab} = \left( \begin{array}{cccc} 0&0&0&0\\ 0&0&0&0 \\ 0&0& \widehat H(\psi,t) &
\;\; \widehat P(\psi,t) \\ 0&0&\widehat P(\psi,t)&-\widehat H(\psi,t) \end{array} \right)
\end{equation}
and Eintsein's linearized equation $\dot R_{ab}=0$ reduces to $\Box_2 \widehat H = \Box_2 \widehat P =0$ \cite{Ishibashi:2011ws}, that is (see 
equation (\ref{Fi2}))
\begin{align} \label{sg}
H &= \widehat H_l (\psi) +  \widehat H_r(\psi + \ln (|t|/v_o^2)) + \tfrac{C_H}{\g} \; \ln (|t|/v_o^2) \\
P &=  \widehat P_l (\psi) +  \widehat P_r(\psi + \ln (|t|/v_o^2)) + \tfrac{C_P}{\g} \; \ln (|t|/v_o^2).
\end{align}
We will   set $C_H=C_P=0$ to keep the perturbation bounded as $t \to -\infty$ (note that $H$ and $H$ are gauge invariant 
fields in the linearized gravity theory \cite{Ishibashi:2011ws}). For the perturbation (\ref{tgp})-(\ref{sg}) we obtain
\begin{equation}
\ddot K = \dot R_{ab}{}^{cd} \dot R^{cd}{}^{ab} = 32 t^{-2}\; [ \widehat H_r'' \; \widehat H_l'' + \widehat P_r'' \; \widehat P_l'' + \widehat H_r'' \widehat H_l' +\widehat P_r'' \widehat P_l' 
- \widehat H_l'' \widehat H_r'-\widehat P_l'' \widehat P_r'-\widehat H_r'\widehat H_l'-\widehat P_r'\widehat P_l'],
\end{equation}
which decays along past oriented causal curves and diverges as the future Cauchy horizon is approached. Once again, this divergence could possibly be canceled from 
$(y,z)$-independent contributions to $\ddot K$ from the $(y,z)$-dependent piece of $\dot g _{ab}$, but this could only happen after fine tuning, and not 
for generic solutions around $\eta_{ab}$

\section{Discussion}

Penrose's  heuristic argument anticipating a curvature singularity at the Cauchy horizon of a Kerr-Newman black hole applies to the 
horizon $h$ of Misner spacetime. This is  readily seen by inspecting Figure \ref{cili}: an observer crossing the horizon 
is exposed to the information traveling -in the geometric optics approximation- along the infinitely many geodesics 
of the form (\ref{t0<0}) originating in his past. He  is thus expected to measure a divergent energy density, as it is easily 
checked, e.g., in our simplest example: the $(y,z)$-independent  scalar field (\ref{fi0}) with $c=0$. The stress-energy-momentum 
tensor of this field is
\begin{equation} \label{semt}
T_{ab} = \left(
 \begin{array}{cccc} 
\hat r ' {}^2 + \hat l ' {}^2 &  \hat r ' {}^2 /t & 0 & 0 \\
 \hat r ' {}^2 /t &  \hat r ' {}^2 /t^2 & 0 & 0 \\ 0 & 0 & 2  \hat r ' {}  \hat l ' {} / t & 0 \\0 & 0 & 0 &
2  \widehat r ' {}  \widehat l ' {} / t 
\end{array} 
\right) 
\end{equation}
where we have suppressed the arguments in  $\hat l '(\psi)$ and $\hat r ' (\psi + \ln (|t|/v_o^2))$ and the order of coordinates 
above is $(\psi,t,y,z)$. An observer crossing the horizon with four-velocity $u=\dot \psi \, \p / \p \psi + \dot t \, \p / \p t +\dot y \, \p / \p y +
\dot z \, \p / \p z$ has $\dot t \neq 0$ at $t =0$. Note that these coordinates are  valid beyond the horizon and hence $\dot \psi, \dot t, \dot y$ and $ \dot z$
 must all 
be finite at $t=0$. The energy density the observer  measures is, after using the condition $u^c u_c$ to eliminate the $\dot \psi \dot t$ term, 
\begin{equation}
\rho = T_{ab} u^a u ^b = \dot \psi {}^2  (  \hat l ' {}^2 - \hat r ' {}^2 ) + \frac{2 \hat r '{}^2}{t} ( 1 + \dot y{}^2 +   \dot z{}^2 )
+ \frac{2 \hat r '{} \hat l{}'}{t} (\dot y{}^2 +   \dot z{}^2 ) + \left( \frac{\hat r{}'}{t} \right)^2 \dot t{}^2.
\end{equation}
Only the first term on the right hand side of above remains finite as $t \to 0^-$, the others all diverge 
except for the trivial $\hat r =0$ case. 
It is interesting to note, however, that there is no curvature singularity in the full Einstein-scalar field theory unless {\em both} 
$\hat r$ and $\hat l$ are different from zero. This is  seen by setting $c=0$ in equation (\ref{divsca}), which gives 
\begin{equation} \label{divsca2}
\ddot{ \R}_{(0)(0)} = \frac{-4\hat l'(\psi) }{\gamma t} \;  \hat r'(\psi + \ln (|t|/v_o^2)).  
\end{equation}
Thus, there are situations 
where the energy density measured by an observer at the horizon diverges while no curvature singularity forms. This happens  
because in  this highly relativistic regime, the pressure/tension cancels the energy density effect on $T^a{}_a \propto R$ unless both left and right 
moving waves are present. This can easily be seen 
 from (\ref{semt}): the trace of the two by two $(\psi,t)$ block vanishes and only the $(y,z)$ tensions/pressures, which 
contain $\hat l{}' \hat r{}'$ products contributes to $T^a{}_a$. Note that this happens without violating energy conditions; 
as is well known, the stress-energy-momentum tensor of a scalar field satisfies the strong as well as 
 the dominant energy conditions. $T^a{}_b$ above can indeed be diagonalized to the form 
$T_a{}^b = 2/t \; \text{diag}(|\hat r ' \hat l '|, -|\hat r ' \hat l '|  ,  \hat r ' \hat l ' ,  \hat r ' \hat l ')$ in a specific orthonormal tetrad.

\acknowledgements This work was partially funded from Grants No. PIP
11220080102479 (Conicet-Argentina), and No. Secyt-UNC 05/B498 (Universidad Nacional de C\'ordoba). Pedro Denaro is a fellow
 of Conicet.

\end{document}